\documentclass[journal,twocolumn,twoside]{IEEEtranTCOM}
\ifCLASSINFOpdf
\else
\fi

\usepackage{epsfig,scalefnt,multirow}
\usepackage{url}
\usepackage{ifthen}
\usepackage{mathtools}

\usepackage{cite}
\usepackage{graphicx}
\usepackage{amssymb}
\usepackage{caption}
\usepackage{booktabs}
\usepackage{amsmath}
\usepackage{epstopdf}
\usepackage{algorithm}
\usepackage{algpseudocode}
\usepackage{algpascal}
\usepackage{cases}
\usepackage{xcolor}
\usepackage{multicol}

\usepackage{stfloats}
  \def\cC{{\mathcal{C}}}

 \def\cN{{\mathcal{N}}}

\def\Re{\mathop{\mathrm{Re}}}
\def\Im{\mathop{\mathrm{Im}}}

\def\b0{{\pmb{0}}} 

\def\ba{{\mathbf{a}}} \def\bb{{\mathbf{b}}}  
 \def\bff{{\mathbf{f}}} \def\bg{{\mathbf{g}}} \def\bh{{\mathbf{h}}}

\def\bu{{\mathbf{u}}}  \def\bw{{\mathbf{w}}} \def\bx{{\mathbf{x}}}
\def\by{{\mathbf{y}}}   

\def\bA{{\mathbf{A}}}  \def\bC{{\mathbf{C}}} 
 \def\bF{{\mathbf{F}}}  \def\bH{{\mathbf{H}}}
\def\bI{{\mathbf{I}}}  \def\bK{{\mathbf{K}}} 
\def\bM{{\mathbf{M}}}   
 \def\bR{{\mathbf{R}}} \def\bS{{\mathbf{S}}} 
  \def\bW{{\mathbf{W}}} 
\def\bY{{\mathbf{Y}}} 

\DeclarePairedDelimiter\norm{\lVert}{\rVert}

\begin{document}
%
\title{Massive MIMO Channel Prediction:\\ Kalman Filtering vs. Machine Learning}
%
%
%

\author{Hwanjin~Kim, Sucheol~Kim, Hyeongtaek~Lee, Chulhee Jang, Yongyun Choi, and~Junil~Choi 
	\thanks{This work was partly supported by the Network Business, Samsung Electronics Co., LTD, by the MSIT(Ministry of Science and ICT), Korea, under the ITRC(Information Technology Research Center) support program(IITP-2020-0-01787) supervised by the IITP(Institute of Information \& Communications Technology Planning \& Evaluation), and by IITP grant funded by the Korea government(MSIT) (No. 2016-0-00123, Development of Integer-Forcing MIMO Transceivers for 5G \& Beyond Mobile Communication Systems).}
	\thanks{H. Kim, S. Kim, H. Lee, and J. Choi are with the School of Electrical Engineering, Korea
		Advanced Institute of Science and Technology, Daejeon 34141, South Korea
		(e-mail: jin0903@kaist.ac.kr; loehcusmik@kaist.ac.kr; htlee8459@kaist.ac.kr; junil@kaist.ac.kr).}
	\thanks{C. Jang and Y. Choi are with the Network Business, Samsung Electronics Co., LTD (e-mail: chulhee.jang@samsumg.com; yongyun.choi@samsung.com).}}

\IEEEpeerreviewmaketitle

\maketitle

\begin{abstract}
	This paper focuses on channel prediction techniques for massive multiple-input multiple-output (MIMO) systems. Previous channel predictors are based on theoretical channel models, which would be deviated from realistic channels. In this paper, we develop and compare a vector Kalman filter (VKF)-based channel predictor and a machine learning (ML)-based channel predictor using the realistic channels from the spatial channel model (SCM), which has been adopted in the 3GPP standard for years. First, we propose a low-complexity mobility estimator based on the spatial average using a large number of antennas in massive MIMO. The mobility estimate can be used to determine the complexity order of developed predictors. The VKF-based channel predictor developed in this paper exploits the autoregressive (AR) parameters estimated from the SCM channels based on the Yule-Walker equations. Then, the ML-based channel predictor using the linear minimum mean square error (LMMSE)-based noise pre-processed data is developed. Numerical results reveal that both channel predictors have substantial gain over the outdated channel in terms of the channel prediction accuracy and data rate. 
	The ML-based predictor has larger overall computational complexity than the VKF-based predictor, but once trained, the operational complexity of ML-based predictor becomes smaller than that of VKF-based predictor. 
\end{abstract}


%
\IEEEpeerreviewmaketitle

\section{Introduction}
%
%
%
%
\IEEEPARstart{A}{ccurate} channel state information (CSI) at base stations (BSs) is essential to fully exploit massive multiple-input multiple-output (MIMO) systems, which are one of the key techniques for 5G and beyond wireless communication systems \cite{Marzetta06}. Wireless channels vary in time due to the mobility of user equipment (UE) in practice \cite{Papa17}, and the CSI at the BS could be outdated, resulting in significant performance degradation in massive MIMO \cite{6608213}. The best way to resolve the outdated CSI problem without additional channel training overhead is to predict future channels based on past CSI \cite{5947055,Kong15}.

The conventional way of channel estimation and prediction in massive MIMO usually relies on the Wiener filtering or Kalman filtering assuming model-based analytical channels \cite{6608213, 6815892, 8724442, 6823657, Choi142, 7938362, Kim2019, 7511214, Haykin96, Baddour05,Kashyap17,Shikur15}. A simple first-order Gauss-Markov process channel was considered in \cite{7511214} while more complex autoregressive (AR) or autoregressive moving average (ARMA) models, which are linear stochastic models describing correlated random processes \cite{Haykin96}, were taken into account in \cite{Baddour05,Kashyap17,Shikur15}. Although effective, these approaches are based on simple analytical models for long-term channel statistics, e.g., rectangular power spectrum to represent the temporal variation of channels in \cite{Kashyap17}, which may not hold in practice.  

Recently, data-driven machine learning (ML)-based channel estimation and prediction methods were proposed for massive MIMO systems in \cite{Wang17, Wen18, Wang19, Dong19, Soltani19, 8815557}. The ML-based approaches can discover inherent linear or nonlinear channel characteristics from sufficient amount of channel data without assuming any prior knowledge of channel model. A deep convolutional neural network (CNN)-based massive MIMO channel estimator using spectral correlation was proposed in \cite{Dong19}, and image super resolution and image restoration networks were exploited in \cite{Soltani19} to estimate communication channels considering the channel as two-dimensional images. For the channel prediction, a CNN-AR based channel predictor by leveraging an auto-correlation function pattern was developed in \cite{8815557}. Most of previous ML-based techniques, however, assumed perfect, noiseless channel data during the training phase, which is impractical.

Both model-based and ML-based channel prediction techniques may suffer from high complexity. It is possible for the BS to predict the channel with minimal complexity once the BS has the knowledge of UE mobility, which determines how fast the channel varies in time. For the UE mobility estimation, maximum likelihood, power spectrum density, and channel covariance were respectively exploited in \cite{Krasny01, Baddour051}, and \cite{Zheng09}. Similar to the channel estimation and prediction, these mobility estimators are based on theoretical channel models, e.g., the Rician fading, which may not be able to accurately represent realistic channels. Moreover, the previous mobility estimators require a large number of time samples to obtain mobility estimates, which hinders their practicality.

Different from previous approaches, we consider the spatial channel model (SCM) \cite{SCM}, which has been adopted in the 3GPP standard for years, to consider realistic wireless channel environments in this paper. We first propose a mobility estimator, dubbed as the spatial average of temporal correlation (SATC)-based mobility estimator, using only a few numbers of time samples. The estimated mobility can be used to balance the complexity and accuracy of channel predictors. Then, we develop and compare the vector Kalman filter (VKF)-based predictor and ML-based predictor for time-varying massive MIMO channels. In the VKF-based prediction, we estimate the vector AR parameters by the Yule-Walker equations \cite{Haykin96} using the sampled auto-correlation matrix and predict the time-varying channels with the VKF. In the ML-based prediction, we use the multi-layer perceptron (MLP) after linear minimum mean square error (LMMSE)-based pre-processing noisy received signals. The numerical results show that the prediction accuracy of VKF-based and ML-based predictors is comparable with respect to the time slot and number of samples. We also compare the complexity of VKF-based and ML-based predictors. It turns out that the total complexity of ML-based predictor is much higher than that of VKF-based predictor. After trained, however, it becomes the opposite, i.e., the ML-based predictor becomes far less complex than the VKF-based predictor. It is also possible for the BS to have more advanced processors, e.g., neural processing units (NPUs), which could even facilitate the effectiveness of ML-based predictors.\footnote{Note that the BS does not need to perform any firmware update to train the ML-based predictor for new environments.} We believe our findings in this paper can guide system operators to select a proper channel predictor depending on their operational environment.

The rest of paper is organized as follows. In Section \ref{Section:system model}, we explain a system model including the SCM and a general framework of the channel prediction. In Section \ref{Section:mobility}, we implement the low-complexity mobility estimator. Then, we develop the VKF-based predictor with the AR parameter estimation in Section \ref{Section:VKF} and explain the ML-based predictor using pre-processed received signals in Section \ref{Section:MLP}. After analyzing the complexity and verifying the numerical results in Section \ref{Section:numerical result}, we conclude the paper in Section \ref{Section:conclusion}.

\textbf{Notation:} Lower case and upper case bold letters represent column vectors and matrices. $\bA^{\mathrm{T}}$, $\bA^{\mathrm{H}}$, $\bA^{\dagger}$, and $\underline{\ba}$ denote the transpose, conjugate transpose, pseudo inverse, and column-wise vectorization of matrix $\bA$. $\mathbb{E}[\cdot]$ represents the expectation, and $\Re(\cdot)$, $\Im(\cdot)$ denote the real part and imaginary part of variable. $\text{vec}(\cdot)$ denotes the column-wise vectorization. ${\boldsymbol{0}}_{m\times n}$ represents the $m \times n$ all zero matrix, ${\boldsymbol{0}}_{m}$ is used for the $m \times m$ all zero matrix, and $\bI_m$ denotes the $m \times m$ identity 
matrix. ${\mathbb{C}}^{m \times n}$ represents the set of all $m \times n$ complex matrices. $|{\cdot}|$ denotes the amplitude of scalar, and $\norm{\cdot}$ represents the $\ell_2$-norm of vector. $\cC \cN(m,\sigma^2)$ denotes the complex normal distribution with mean $m$ and variance $\sigma^2$. $\by_1^n$ represents the vector sequences $\{\by_1,\cdots,\by_n\}$. $\mathcal{O}$ denotes the Big-O notation.

\section{System model and general framework}\label{Section:system model}

\subsection{System model}\label{subection:system model}
\begin{figure}[tbp]
	\centering
	\includegraphics[width=8 cm]{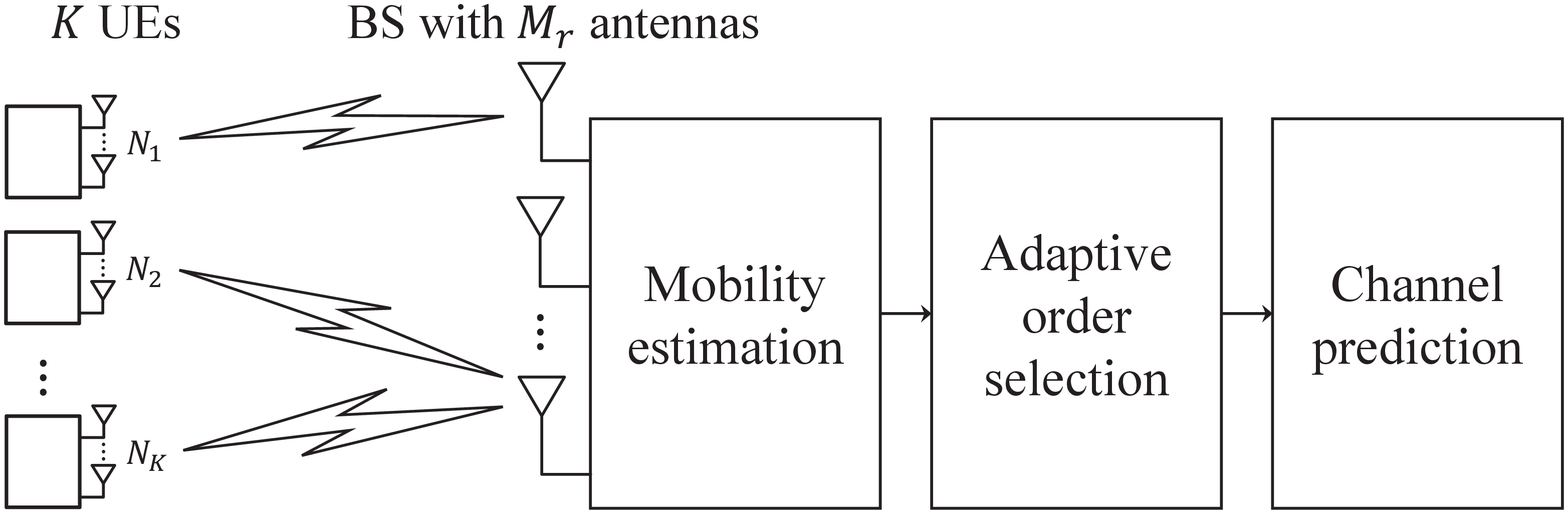}
	\caption{Massive MIMO systems with a BS with $M_r$ antennas and $K$ UEs with $N_k$ antennas each. The BS predicts the channel using proper complexity order based on the mobility estimate.}
	\label{figure1}
\end{figure}

We consider an uplink narrow-band single-cell massive MIMO system as in Fig. \ref{figure1} with a BS with $M_r$ antennas, $K$ UEs with $N_k$ antennas each, making the total number of transmit antennas at the UEs $N_t=\sum_{k=1}^K N_k$. The received signal at the $n$-th time slot is given by
\begin{align}
\by_n&=\sqrt{\rho}\bar{\bH}_n\bx_n+\bw_n\notag\\
&=\sqrt{\rho}\bH_{n,k}\bx_{n,k}+\sqrt{\rho}\sum_{j \neq k}\bH_{n,j}\bx_{n,j}+\bw_n,\label{MIMO}
\end{align}
where $\rho$ is the signal-to-noise ratio (SNR), $\bar{\bH}_n=\begin{bmatrix} \bH_{n,1} &\cdots &\bH_{n,K}\end{bmatrix} \in \mathbb{C}^{M_r \times N_t}$ is the overall channel matrix, $\bH_{n,k}=\begin{bmatrix} \bh_{n,k,1} &\cdots &\bh_{n,k,N_k}\end{bmatrix}\in \mathbb{C}^{M_r \times N_k}$ is the channel between the $k$-th UE and BS, ${\bx_n=\begin{bmatrix} \bx_{n,1}^\mathrm{T} &\cdots &\bx_{n,K}^\mathrm{T}\end{bmatrix}^\mathrm{T} \in \mathbb{C}^{N_t \times 1}}$ is the transmitted symbol vector from all UEs, and $\bw_n\sim \mathcal{C}\mathcal{N}({\boldsymbol{0}}_{M_r\times 1},\bI_{M_r})$ is the complex Gaussian noise. 

To reduce the interference, we use the zero-forcing (ZF) combiner
\begin{align}
	\tilde{\bF}_{n}^\mathrm{T}=\left(\hat{\bar{\bH}}_{n}^\mathrm{H}\hat{\bar{\bH}}_{n}\right)^{-1}\hat{\bar{\bH}}_{n}^\mathrm{H},
	\end{align}where $\hat{\bar{\bH}}_{n}$ is the predicted channel matrix. We set the receive combiner $\bar{\bF}_{n}=\begin{bmatrix}\bF_{n,1} &\cdots &\bF_{n,K}\end{bmatrix}$ based on $\tilde{\bF}_{n}$ to satisfy the unit-norm constraint. Note that $\bF_{n,k}=\begin{bmatrix}\bff_{n,k,1} &\cdots &\bff_{n,k,N_k}\end{bmatrix}$ represents the receive combiner for the $k$-th UE, satisfying $\norm{\bff_{n,k,m}}^2=1$ for all $1\leq m \leq N_k$ and $1\leq k \leq K$. 

After applying the receive combiner $\bar{\bF}_{n}$, which is based on the predicted channels, the achievable rate of the $k$-th UE assuming the Gaussian channel inputs is given as
\begin{align}
R_k=\mathbb{E}\left[\sum_{m=1}^{N_k}\log_2{\left(1+\frac{\textrm{S}_{n,k,m}}{\textrm{IUI}_{n,k,m}+\textrm{ISI}_{n,k,m}+\textrm{N}_{n,k,m}}\right)}\right],
\end{align}
where 
\begin{align}
\textrm{S}_{n,k,m}&=\rho\big|\bff_{n,k,m}^\mathrm{T}\bh_{n,k,m}\big|^2, \notag \\
\textrm{IUI}_{n,k,m}&=\rho\sum_{j \neq k}\sum_{m=1}^{N_k}\big|\bff_{n,k,m}^\mathrm{T}\bh_{n,j,m}\big|^2, \notag \\
\textrm{ISI}_{n,k,m}&=\rho\sum_{m'\neq m}\big|\bff_{n,k,m}^\mathrm{T}\bh_{n,k,m'}\big|^2, \notag \\
\textrm{N}_{n,k,m}&=\big|\bff_{n,k,m}^\mathrm{T}\bw_n\big|^2.
\end{align}The achievable sum-rate is then defined as
\begin{align}
R=\sum_{k=1}^K R_k. \label{achivable sum-rate}
\end{align}

Since channel prediction techniques can be applied to each UE separately, we focus on a single UE case and drop the UE index $k$ throughout the paper. The BS predicts the channel after receiving length $\tau$ pilot sequences from the UE as
\begin{align}
\bY_n=\sqrt{\rho}\bH_n \boldsymbol{\Psi}_n^\mathrm{T}+\bW_n, \label{measurement}
\end{align}where $\bY_n \in \mathbb{C}^{M_r\times \tau}$ is the received signal, $\boldsymbol{\Psi}_n\in \mathbb{C}^{\tau \times N}$ is the pilot matrix assuming that the pilot sequences are column-wise orthogonal, i.e., $\boldsymbol{\Psi}_n^\mathrm{T}\boldsymbol{\Psi}_n^*=\tau\bI_N$, and $\bW_n\in \mathbb{C}^{M_r\times \tau}$ is the Gaussian noise. 
For the sake of simplicity, the received signal is vectorized as
\begin{align}
\underline{\by}_n=\bar{\boldsymbol{\Psi}}_n \underline{\bh}_n +\underline{\bw}_n, \label{measurement2}
\end{align}
where $\bar{\boldsymbol{\Psi}}_n=(\sqrt{\rho}\boldsymbol{\Psi}_n\otimes \bI_{M_r})\in \mathbb{C}^{M_r\tau\times M_r N}$, $\underline{\bh}_n=\textrm{vec}(\bH_n)\in \mathbb{C}^{M_rN\times 1}$, and $\underline{\bw}_n=\text{vec}(\bW_n)\in \mathbb{C}^{M_r \tau\times 1}$. 

To predict $\bH_n$, most previous works assumed certain analytical models to represent $\bH_n$. The SCM channels considered in this paper can also be represented with stochastic parameters. As in \cite{Pan07}, the SCM channel from the $u$-th UE antenna to the $m$-th BS antenna through the $t$-th path at the $n$-th time slot can be written as
\begin{align}
	h_{u,m,t,n}^\mathrm{SCM}=\sqrt{\frac{P_t}{L_s}}\sum_{l=1}^{L_s}\Big{\{}&\exp\left(jkd_m \sin(\theta_{t,l,\text{AoA}})\right)\exp(j\phi_{t,l}) \notag \\ 
	\cdot &\exp(jkd_u\sin(\theta_{t,l,\text{AoA}})) \notag \\
\cdot&\exp\left(jk{|v|}\cos(\theta_{t,l,\text{AoD}}-\theta_v)n\right)\Big{\}},\label{SCM}
\end{align}
where $P_t$ is the power of the $t$-th path, $L_s$ is the number of subpaths per-path, $j=\sqrt{-1}$, $k=\frac{2\pi}{\lambda}$ is the wave number with $\lambda$ denoting the carrier wavelength, $d_m$ is the distance of BS antenna element $m$ from the reference antenna $m=1$, $d_u$ is the distance of UE antenna element $u$ from the reference antenna $u=1$, $\theta_{t,l,\textrm{AoA}}$ is the angle-of-arrival (AoA) for the $l$-th subpath of the $t$-th path at the BS, $\theta_{t,l,\textrm{AoD}}$ is the angle-of-departure (AoD) for the $l$-th subpath of the $t$-th path at the UE, $\phi_{t,l}$ is the phase of the $l$-th subpath of the $t$-th path, $|v|$ is the magnitude of UE mobility, and $\theta_v$ is the angle of moving direction of UE. Then, the $M_r\times N$ channel matrix $\bH_n$ is given as
\begin{align}
\bH_n=\begin{bmatrix}\sum_t h_{1,1,t,n}^\mathrm{SCM} &\cdots &\sum_t h_{N,1,t,n}^\mathrm{SCM}\\
\vdots &\ddots &\vdots \\
\sum_t h_{1,M_r,t,n}^\mathrm{SCM} &\cdots &\sum_t h_{N,M_r,t,n}^\mathrm{SCM}
\end{bmatrix}.
\end{align}

Note that the SCM incorporates the important parameters in the wireless environment channels as in (\ref{SCM}). Thus, we believe it is critical to evaluate developed predictors with the SCM to guarantee their usefulness in practice. It is difficult, however, to predict all the SCM parameters directly due to its complex structure. Note that the parameters in (\ref{SCM}) are temporally correlated random processes. Thus, it would be possible to predict the channel itself, i.e., $\bH_n$, not individual parameters, using the temporal correlation inherent in the channels.

\subsection{General framework of channel prediction}
As in Fig. \ref{figure1}, the BS first estimates the mobility of UE, which can be done with very low overhead in time by using a large number of antennas in massive MIMO. Using the mobility estimate, the BS then determines proper complexity order for the channel prediction to balance the prediction complexity and accuracy.\footnote{The channel prediction with adaptive complexity according to the mobility for single-input single-output (SISO) SCM channels was proposed in \cite{6632104}.} With the proper complexity order, the BS predicts the channel $\underline{\bh}_{n+1}$ with the previous measurements $\big\{\underline{\by}_{n-n_o+1}, \cdots, \underline{\by}_n\big\}$ for $n_o \geq 1$ where $n_o$ is the complexity order. Formally, we can set an optimization problem
\begin{alignat}{3}
&\text{minimize} &&~{\left\|\underline{\bh}_{n+1}-\hat{\underline{\bh}}_{n+1}\right\|}^2,\\
&\text{subject to  }&&~{\hat{\underline{\bh}}}_{n+1}=f\big(\underline{\by}_{n-n_o+1},\cdots,\underline{\by}_{n}\big),\\
& &&~n_o=g(\hat{v}),\label{optimization}
\end{alignat} 
where $\underline{\hat{\bh}}_{n+1}$ is the predicted channel, $f(\cdot)$ is an arbitrary predictor, and $g(\cdot)$ is the relation between the complexity order of predictor $n_o$ and estimated mobility of UE $\hat{v}$. Due to the complex structure of realistic SCM channel, the optimization problem is highly nonlinear. Therefore, we develop the two channel predictors with the tractable complexity based on the Kalman filtering and machine learning.

\begin{figure}[t]
	\centering
	\includegraphics[width=9.4 cm]{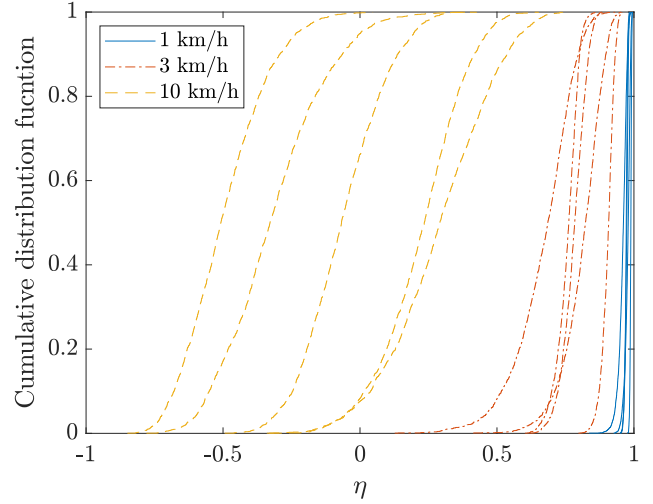}
	\caption{CDF of $\eta$ with different UE geometries and mobilities.}
	\label{figure2}
\end{figure}

\section{Mobility estimation}\label{Section:mobility}
There is in general trade-off relation between channel prediction performance and complexity. If the BS has the mobility information of UE, which is the main factor that determines how ``fast'' the channel between the UE and BS varies, it would be possible to balance between the prediction performance and complexity. Most of previous mobility estimation methods are based on simple channel models and require a large number of channel snapshots in time, making them impractical \cite{Krasny01, Baddour051, Zheng09}. 

In this section, we propose the spatial average of temporal correlation (SATC)-based mobility estimator, which works well in time-varying channels. The proposed mobility estimator requires just two channel snapshots in time by spatially averaging the temporal correlation in massive MIMO. Precisely, the BS first obtains
\begin{align}
\eta=\Re\left(\frac{\underline{\bh}_{n-1}^\mathrm{H}\underline{\bh}_{n}}{\norm{\underline{\bh}_{n-1}}\norm{\underline{\bh}_{n}}}\right),\label{SATC}
\end{align}and compare $\eta$ with given thresholds to estimate the mobility of UE. We only take the real part in (\ref{SATC}) since $\eta$ will be close to one when the UE moves slowly, and the channel does not vary much. For the sake of simplicity, we assume the perfect channel in the SATC-based estimator. In general, the minimum mean square error (MMSE) estimate or least square (LS) estimates can be used to obtain $\eta$.

In Fig. \ref{figure2}, we plot the cumulative distribution function (CDF) of $\eta$, i.e., $F_X(\eta)=\mathrm{Pr}\{X<\eta\}$ with different UEs experiencing various mobilities. We set different geometries for the UEs with the same mobility. Fig. \ref{figure2} shows that the BS can estimate the UE mobilities, even when the UEs experience different geometries, by only using two channel snapshots. Using the estimated mobility, we discuss how to set the complexity order of developed channel predictors in Section \ref{Section:numerical result}. It requires further efforts to accurately estimate the Doppler frequency, which is out of scope of this paper.

\section{Kalman filter-based prediction}\label{Section:VKF}
In this section, we first elaborate how to estimate the AR parameters from the SCM channels.\footnote{The AR model has a computational advantage over the ARMA model. When estimating the parameters, the AR model requires only linear equations whereas the ARMA model needs to solve nonlinear equations. Besides, based on an analytical channel model, it was shown in \cite{Kashyap17} that the Kalman filtering using the AR and ARMA models have almost the same channel prediction performance. Therefore, we use the AR model instead of the ARMA model to represent the SCM channels.} Based on the established AR model and estimated parameters, we develop the VKF-based predictor to predict the SCM channel variation in time.

\subsection{AR parameter estimation}
The AR models are widely used in stochastic processes \cite{Haykin96}. The vector AR($p$) model is given by
\begin{align}
\underline{\bh}_n=\sum_{i=1}^p\boldsymbol{\Phi}_i \underline{\bh}_{n-i} +\bu_n,\label{AR}
\end{align}where $p$ is the AR-order, $\boldsymbol{\Phi}_i\in\mathbb{C}^{M_r N \times M_r N}$ is the $i$-th AR parameter matrix, and $\bu_n\sim \mathcal{C}\mathcal{N}(\boldsymbol{0}_{M_r N\times 1},\boldsymbol{\Sigma})$ is the innovation process. 
We estimate the two sets of AR model parameters, which are the AR parameter matrix $\boldsymbol{\Phi}_i$ and  covariance matrix $\boldsymbol{\Sigma}$ of innovation process $\bu_n$ via the Yule-Walker equations, in order to uniquely define the vector AR($p$) model \cite{Haykin96}. The Yule-Walker equations can be represented as a matrix form
\begin{align}
{\begin{bmatrix} \bR(1) &\bR(2) &\cdots &\bR(p) \end{bmatrix}=
	\begin{bmatrix}\boldsymbol{\Phi}_1 &\boldsymbol{\Phi}_2 &\cdots &\boldsymbol{\Phi}_p\end{bmatrix}\bar{\bR},}\label{YW equation}
\end{align}
with
\begin{align}
\bar{\bR}=\begin{bmatrix} \bR(0) &\bR(1) &\cdots &\bR(p-1) \\
\bR^\mathrm{H}(1) &\bR(0) &\cdots &\bR(p-2)\\
\vdots &\vdots &\ddots &\vdots \\
\bR^\mathrm{H}(p-1) &\bR^\mathrm{H}(p-2) &\cdots &\bR(0)
\end{bmatrix},
\end{align}
where $\bR(i)=\mathbb{E}\left[{\underline{\bh}_n\underline{\bh}_{n-i}^\mathrm{H}}\right]$ is the auto-correlation matrix of $\underline{\bh}_n$ with the lag $i$. Thus, we can obtain the AR parameters by solving (\ref{YW equation}) as
\begin{align}
\begin{bmatrix}{\boldsymbol{\Phi}}_1 &{\boldsymbol{\Phi}}_2 &\cdots &{\boldsymbol{\Phi}}_p\end{bmatrix}=\begin{bmatrix} \bR(1) &\bR(2) &\cdots &\bR(p) \end{bmatrix}\bar{\bR}^{-1},\label{solve YW}
\end{align}where the {covariance} matrix $\boldsymbol{\Sigma}$ is given by
\begin{align}
\boldsymbol{\Sigma}=\bR(0)-\sum_{i=1}^p\boldsymbol{\Phi}_i\bR^\mathrm{H}(i).
\end{align}
To avoid the large matrix inversion in the Yule-Walker equations, the Levinson-Durbin recursion can be used \cite{durbin60}.

\begin{algorithm}[tbp]
	\renewcommand{\arraystretch}{1.3}
	\begin{algorithmic}[1]
		\caption{Kalman Filter-Based Channel Predictor}
		\State Initialization:
		\begin{align*}
		\underline{\hat{\tilde{\bh}}}_{0|0}&=\boldsymbol{0}_{M_r N p \times 1},\\
		\bM_{0|0}&=\mathbb{E}\left[\underline{\tilde{\bh}}_0\underline{\tilde{\bh}}_0^\mathrm{H}\right]=\bar{\bR}
		\end{align*}
		\State Prediction:
		\begin{align*}
		\underline{\hat{\tilde{\bh}}}_{n+1|n}&=\mathbb{E}\left[\underline{\tilde{\bh}}_{n+1}\big|\underline{\by}_1^{n}\right]\notag\\
		&=\bar{\boldsymbol{\Phi}}\underline{\hat{\tilde{\bh}}}_{n|n} 
		\end{align*}
		\State Minimum prediction MSE matrix:
		\begin{align*}
		\bM_{n+1|n}&=\mathbb{E}\left[\left(\underline{\tilde{\bh}}_{n+1}-\underline{\hat{\tilde{\bh}}}_{n+1|n}\right)\left(\underline{\tilde{\bh}}_{n+1}-\underline{\hat{\tilde{\bh}}}_{n+1|n}\right)^\mathrm{H}\big|\underline{\by}_1^{n}\right]\\
		&=\bar{\boldsymbol{\Phi}}\bM_{n|n}\bar{\boldsymbol{\Phi}}^\mathrm{H}+\bar{\boldsymbol{\Theta}}\boldsymbol{\Sigma}\bar{\boldsymbol{\Theta}}^\mathrm{H}
		\end{align*}
		\State Kalman gain matrix:
		\begin{align*}
		\bK_{n+1}=\bM_{n+1|n}\bS^\mathrm{H}\left(\bS\bM_{n+1|n}\bS^\mathrm{H}+\bI_{M_r \tau}\right)^{-1}
		\end{align*}
		\State Correction:
		\begin{align*}
		\underline{\hat{\tilde{\bh}}}_{n+1|n+1}=\underline{\hat{\tilde{\bh}}}_{n+1|n}+\bK_{n+1}\left(\underline{\by}_{n+1}-\bS\underline{\hat{\tilde{\bh}}}_{n+1|n}\right)
		\end{align*}
		\State Minimum MSE matrix:
		\begin{align*}
		\bM_{n+1|n+1}&=\mathbb{E}~\bigg[\left(\underline{\tilde{\bh}}_{n+1}-\underline{\hat{\tilde{\bh}}}_{n+1|n+1}\right)\notag\\
		&~\qquad\times\left(\underline{\tilde{\bh}}_{n+1}-\underline{\hat{\tilde{\bh}}}_{n+1|n+1}\right)^\mathrm{H}\big|\underline{\by}_1^{n+1}\bigg]\notag\\
		&=\left(\bI_{M_r Np}-\bK_{n+1}\bS\right)\bM_{n+1|n}
		\end{align*}
	\end{algorithmic}
\end{algorithm}  

\begin{figure*}[t]
	\centering
	\includegraphics[width=13.5 cm]{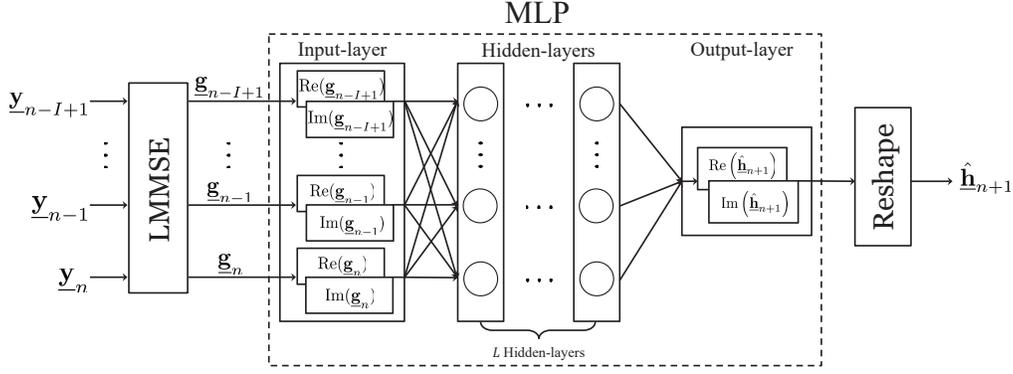}%
	\caption{MLP structure with LMMSE pre-processing.}
	\label{figure3}
\end{figure*}

$\textbf{Remark 1}$: Note that the aggregated auto-correlation matrix $\bar{\bR}$ is required in the Yule-Walker equations. Assuming the channel statistics do not change much in several coherence time intervals, the BS can obtain a sampled auto-correlation matrix $\hat{\bar{\bR}}$ from the measurement vectors $\underline{\tilde{\by}}_n=\begin{bmatrix}\underline{\by}_n^\mathrm{T} &\cdots &\underline{\by}_{n-p+1}^\mathrm{T}\end{bmatrix}^\mathrm{T}$ as
\begin{align}
\hat{\bar{\bR}}=\tilde{\boldsymbol{\Psi}}_n^\dagger\left(\frac{1}{N_s}\sum_{n=1}^{N_s}\underline{\tilde{\by}}_n\underline{\tilde{\by}}_n^\mathrm{H}-\bI_{M_r \tau p}\right)\left(\tilde{\boldsymbol{\Psi}}_n^\mathrm{H}\right)^\dagger,\label{N_s}
\end{align}with 
\begin{align}
\tilde{\boldsymbol{\Psi}}_n=
\begin{bmatrix}\boldsymbol{\bar{\Psi}}_n &\boldsymbol{0}_{M_r\tau \times M_rN} &\cdots &\boldsymbol{0}_{M_r\tau \times M_rN}\\
\boldsymbol{0}_{M_r\tau \times M_rN} &\boldsymbol{\bar{\Psi}}_{n-1} &\cdots &\boldsymbol{0}_{M_r\tau \times M_rN}\\
\vdots &\vdots &\ddots  &\vdots\\
\boldsymbol{0}_{M_r\tau \times M_rN} &\boldsymbol{0}_{M_r\tau \times M_rN} &\cdots &\boldsymbol{\bar{\Psi}}_{n-p+1}
\end{bmatrix},
\end{align}where $N_s$ is the number of measurement vectors. Note that the SNR $\rho$ in $\boldsymbol{\bar{\Psi}}_n$ can be estimated by measuring the received signal power \cite{Hung10}.

$\textbf{Remark 2}$: To have an accurate solution in (\ref{YW equation}) by using $\hat{\bar{\bR}}$ instead of $\bar{\bR}$, the auto-correlation matrix $\hat{\bar{\bR}}$ needs to be non-singular. If the order $p$ is high, however, $\hat{\bar{\bR}}$ could be ill-conditioned. We can resolve this issue using
\begin{align}
\hat{\bar{\bR}}_{\epsilon}=\hat{\bar{\bR}}+\epsilon \bI_{M_rN p},
\end{align}
instead of $\hat{\bar{\bR}}$ where $\epsilon$ is a very small number \cite{Baddour05}. Note that we perform all the simulations in Section \ref{Section:numerical result} with $\hat{\bar{\bR}}_{\epsilon}$.

\subsection{Kalman filter-based prediction}\label{Subsection:Kalman}
In the Kalman filtering, we need to define the state equation and measurement equation to predict the channel sequentially. For convenience, we can express the state equation by rewriting the vector AR($p$) in (\ref{AR}) as an equivalent vector AR model of order 1 as
\begin{align}
\underline{\tilde{\bh}}_n=\bar{\boldsymbol{\Phi}}\underline{\tilde{\bh}}_{n-1}+\bar{\boldsymbol{\Theta}}\bu_n,
\end{align}where $\underline{\tilde{\bh}}_n=\begin{bmatrix}\underline{\bh}_n^\mathrm{T} &\cdots &\underline{\bh}_{n-p+1}^\mathrm{T}\end{bmatrix}^\mathrm{T}\in\mathbb{C}^{M_r Np\times 1}$ is the state vector at the $n$-th time slot with the order $p$, $\bu_n$ is the innovation process in (\ref{AR}), and the state transition matrices $\bar{\boldsymbol{\Phi}}$ and $\bar{\boldsymbol{\Theta}}$ are \begin{align}
	&\bar{\boldsymbol{\Phi}}=\notag\\
	&\begin{bmatrix}\boldsymbol{\Phi}_1 &\boldsymbol{\Phi}_2 &\cdots &\boldsymbol{\Phi}_{p-1} &\boldsymbol{\Phi}_p\\
	\bI_{M_r N} &\boldsymbol{0}_{M_r N} &\cdots &\boldsymbol{0}_{M_r N} &\boldsymbol{0}_{M_r N}\\
	\boldsymbol{0}_{M_r N} &\bI_{M_r N} &\cdots &\boldsymbol{0}_{M_r N } &\boldsymbol{0}_{M_r N}\\
	\vdots &\vdots &\ddots &\vdots &\vdots\\
	\boldsymbol{0}_{M_r N} &\boldsymbol{0}_{M_r N} &\cdots &\bI_{M_r N} &\boldsymbol{0}_{M_r N}
	\end{bmatrix}\in\mathbb{C}^{M_rNp \times M_rNp}, \label{Phi}
	\end{align}
\begin{align}
\bar{\boldsymbol{\Theta}}=\begin{bmatrix}\bI_{M_r N} \\ \boldsymbol{0}_{M_r N} \\ \vdots\\ \boldsymbol{0}_{M_r N}\end{bmatrix}\in \mathbb{C}^{M_r Np \times M_r N}.
\end{align}The measurement equation can be reformulated from (\ref{measurement2}) as
\begin{align}
\underline{\by}_n=\bS \underline{\tilde{\bh}}_n+\underline{\bw}_n,
\end{align}
where $\bS=\begin{bmatrix}\bar{\boldsymbol{\Psi}}_n &\boldsymbol{0}_{M_r \tau \times M_r N} &\cdots &\boldsymbol{0}_{M_r \tau \times M_r N}\end{bmatrix} \in\mathbb{C}^{M_r \tau \times M_r Np}$ is the measurement matrix, and $\underline{\bw}_n$ is the Gaussian noise in (\ref{measurement2}). Based on these parameters and equations, the VKF-based prediction method is summarized in Algorithm 1. The predicted channel $\hat{\underline{\bh}}_{n+1|n}$ is the first $M_rN$ elements of $\underline{\hat{\tilde{\bh}}}_{n+1|n}$ defined in Step 2 of Algorithm 1.

\section{Machine learning-based prediction}\label{Section:MLP}
In this section, we develop the ML-based predictor for the SCM channel. First, we explain the MLP, a popular neural network (NN) structure adopted in this paper. Then, we propose the LMMSE-based received signal pre-processing technique. Finally, we present the training method for the MLP.

\subsection{MLP structure}
While any NN structures can be used for channel prediction, we adopt the MLP structure since it is simple to implement and has good prediction performance as shown in Section \ref{Section:numerical result}. In Fig. \ref{figure3}, the MLP structure consists of input-layer, output-layer, and hidden-layers where the hidden-layer is composed with $L$ fully-connected layers. For the MLP input, we perform the LMMSE pre-processing, as explained in the next subsection, to make the predictor more robust to the noise $\underline{\bw}_n$.

\subsection{LMMSE-based noise pre-processing}
We first define 
\begin{align}
\bC_{\underline{\bh}_n\underline{\by}_n}&=\mathbb{E}\left[\underline{\bh}_n\underline{\by}_n^\mathrm{H}\right],\\
\bC_{\underline{\by}_n}&=\mathbb{E}\left[\underline{\by}_n\underline{\by}_n^\mathrm{H}\right],\\
\bC_{\underline{\bh}_n}&=\mathbb{E}\left[\underline{\bh}_n\underline{\bh}_n^\mathrm{H}\right].
\end{align}Then, $\underline{\bg}_n$, the LMMSE estimate of $\underline{\by}_n$, is given as
\begin{align}
\underline{\bg}_n&=\bC_{\underline{\bh}_n\underline{\by}_n}\bC_{\underline{\by}_n}^{-1}\underline{\by}_n \notag\\ 
&=\bC_{\underline{\bh}_n}\boldsymbol{\bar{\Psi}}_n^\mathrm{H}\left(\boldsymbol{\bar{\Psi}}_n\bC_{\underline{\bh}_n}\boldsymbol{\bar{\Psi}}_n^\mathrm{H}+\bI_{M_r\tau}\right)^{-1}\underline{\by}_n.\label{g_n}
\end{align}As in Section \ref{Section:VKF}, the BS can obtain the sampled auto-covariance $\hat{\bC}_{\underline{\by}_n}$,
\begin{align}
\hat{\bC}_{\underline{\by}_n}=\frac{1}{N_s}\sum_{i=1}^{N_s}{\underline{\by}_i\underline{\by}_i^\mathrm{H}},
\end{align}where $N_s$ is the number of samples. To obtain the sampled auto-covariance $\hat{\bC}_{\underline{\bh}_n}$, we exploit the relation
\begin{align}
{\bC}_{\underline{\by}_n}&=\mathbb{E}\left[\underline{\by}_n\underline{\by}_n^\mathrm{H}\right]\notag\\
&=\mathbb{E}\left[\left(\boldsymbol{\bar{\Psi}}_n\underline{\bh}_n+\underline{\bw}_n\right)\left(\boldsymbol{\bar{\Psi}}_n\underline{\bh}_n+\underline{\bw}_n\right)^\mathrm{H}\right]\notag\\
&=\boldsymbol{\bar{\Psi}}_n\bC_{\underline{\bh}_n}\boldsymbol{\bar{\Psi}}_n^\mathrm{H}+\bI_{M_r\tau}.\label{relation}
\end{align}From (\ref{relation}), we have $\hat{\bC}_{\underline{\bh}_n}=\boldsymbol{\bar{\Psi}}_n^{\dagger}\left(\hat{\bC}_{\underline{\by}_n}-\bI_{M_r \tau}\right)\left(\boldsymbol{\bar{\Psi}}_n^\mathrm{H}\right)^{\dagger}$. Using $\hat{\bC}_{\underline{\by}_n}$, $\hat{\bC}_{\underline{\bh}_n}$, and the received signal $\underline{\by}_n$, the BS can acquire $\underline{\bg}_n$.

By denoting the predicted channel $\hat{\underline{\bh}}_{n+1}$ as the output, the input-output relationship of MLP is given as
\begin{align}
\hat{\underline{\bh}}_{n+1}=f_\Pi\big(\underline{\bg}_{n-I+1},\cdots,\underline{\bg}_{n}\big), \label{MLP_io}
\end{align}where $\Pi$ is the parameter set of MLP, and $I$ is the input-order, which can balance between the MLP complexity and prediction performance. Note that denoising the input data for channel estimation has been proposed in recent works, \cite{8353153, 8651830}; however, these works relied on the deep CNN-based architectures with considerable complexity. On the contrary, the proposed LMMSE-based pre-processing is simple yet practical.

\subsection{MLP training}
In the MLP training phase, the inputs to the MLP are the noise pre-processed channel vectors  $\big\{\underline{\bg}_{n-I+1},\cdots,\underline{\bg}_{n}\big\}$, and the output is the predicted channel vector at the $(n+1)$-th time slot $\hat{\underline{\bh}}_{n+1}$.
For the real-valued MLP architecture, we reshape the inputs to a $2IM_rN$-dimension input-layer, which is the real and imaginary parts of input vectors, i.e., $\left\{\mathrm{Re}\big(\underline{\bg}_{n-I+1}\big), \mathrm{Im}\big(\underline{\bg}_{n-I+1}\big), \cdots, \mathrm{Re}\big(\underline{\bg}_{n}\big), \mathrm{Im}\big(\underline{\bg}_{n}\big)\right\}$. In the hidden-layer, we use $L$ fully-connected layers with $f_l$ nodes for $1 \leq l\leq L$. The output-layer is designed to have $2M_rN$-dimension, which corresponds to the real and imaginary parts of channel vector at the $(n+1)$-th time slot $\left\{\mathrm{Re}\big(\hat{\underline{\bh}}_{n+1}\big), \mathrm{Im}\big(\hat{\underline{\bh}}_{n+1}\big)\right\}$. The last reshape layer combines the real and imaginary parts to reconstruct the complex-valued predicted channel vector $\hat{\underline{\bh}}_{n+1}$.

We use the adaptive moment estimation (ADAM) as the optimizer, and the loss function for the NN training by the mean square error (MSE) between the predicted channel and the noise pre-processed channel ${\underline{\bg}}_{n+1}$, not the true channel $\underline{\bh}_{n+1}$,
\begin{align}
\mathcal{C}_{\text{loss}}=\frac{1}{N_\text{train}}\sum_{n=I}^{N_\text{train}}{\left\|\hat{\underline{\bh}}_{n+1}-\underline{\bg}_{n+1}\right\|}^2,\label{N_train}
\end{align}where $N_\text{train}$ is the number of training samples. Although previous works on channel estimation and prediction using NN, e.g., \cite{Dong19,Soltani19,8815557}, used the true channel for the loss function, this is not possible in practice. Therefore, we have used the pre-processed data $\underline{\bg}_n$ for the NN training and also for the prediction.

\begin{table*}[t!]
	\centering
	\renewcommand{\arraystretch}{1.55}
	\captionsetup{justification=centering}
	\captionsetup{labelsep=newline}
	\caption{Complexity of VKF-based predictor and MLP-based predictor} 
	\resizebox{\linewidth}{!}{%
		\begin{tabular}{l l l l}
			\toprule
			\text{Channel predictor} & Method & \text{Complexity} & \text{Total complexity} \\
			\hline
			\multirow{2}{*}{VKF} & \multicolumn{1}{l}{AR estimation} &\multicolumn{1}{l}{$\mathcal{O}\left((pM_rN)^3\right)$} & \multirow{2}{*}{$\mathcal{O}\left(\left(p^3+1\right)(M_rN)^3\right)$} \\ 
			& \multicolumn{1}{l}{Kalman filtering} & \multicolumn{1}{l}{$\mathcal{O}\left((M_rN)^3\right)$} & \\ \hline
			\multirow{3}{*}{MLP}& \multicolumn{1}{l}{LMMSE estimation} & \multicolumn{1}{l}{$\mathcal{O}\left((M_rN)^3\right)$} & \multirow{3}{*}{$\mathcal{O}\left((N_\text{epoch}N_\text{train}+1)\alpha(I+(L-1)\alpha+1)(M_rN)^2+(M_rN)^3\right)$} \\ 
			& \multicolumn{1}{l}{MLP train} & \multicolumn{1}{l}{$\mathcal{O}\left(N_\text{epoch}N_\text{train}\alpha(I+(L-1)\alpha+1) (M_rN)^2\right)$} & \\ 
			& \multicolumn{1}{l}{MLP test} & \multicolumn{1}{l}{$\mathcal{O}\left(\alpha(I+(L-1)\alpha+1) (M_rN)^2\right)$} & \\ 
			\bottomrule
	\end{tabular}}
	{\label{table1}}
\end{table*}

\section{Complexity analysis and numerical results}\label{Section:numerical result}
In this section, we first analyze the computational complexity of developed predictors. Then, through numerical studies, we determine proper parameter values for the VKF-based and MLP-based predictors. Using these parameter values, we thoroughly compare the two predictors in terms of the prediction accuracy and achievable sum-rate.

\subsection{Complexity analysis}
In this subsection, we analyze the computational complexity of VKF-based predictor with the AR parameter estimation and that of MLP-based predictor based on the LMMSE pre-processing. We use the number of floating-point operations (FLOPs) as the performance metric with the Big-O notation \cite{hunger2005floating}. 

In the VKF-based predictor, we first consider the AR parameter estimation complexity. To estimate the AR parameters in (\ref{YW equation}) for the Kalman filtering, the AR parameter estimation has the complexity of $\mathcal{O}\left((pM_rN)^3\right)$ because of the matrix inversion in (\ref{solve YW}). Also, the Kalman filtering to predict the channel has the complexity of $\mathcal{O}\left((M_rN)^3\right)$ due to the matrix inversion of Kalman gain matrix. Thus, the total complexity of VKF-based predictor is
\begin{align}
C_\text{VKF}=\mathcal{O}\left(\left(p^3+1\right)(M_rN)^3\right).\label{complexity_VKF}
\end{align}

The MLP-based predictor has two types of complexity, i.e., the complexity of training phase and prediction phase, which is usually called as the test phase in machine learning literature. With the number of epochs $N_\text{epoch}$, the number of training samples $N_\text{train}$, and the number of hidden-layer $L$ with $f_l$ nodes, the complexity of training phase becomes \cite{939044}
\begin{align}
&C_\text{train}\notag\\
&=\mathcal{O}\left(N_\text{epoch}N_\text{train}\left(IM_rN f_1+\sum_{l=1}^{L-1}f_l f_{l+1}+f_L M_rN\right)\right)\notag\\
&\stackrel{(a)}{=}\mathcal{O}\big(N_\text{epoch} N_\text{train}\big(\alpha I(M_rN)^2+(L-1)\alpha^2(M_rN)^2\notag\\
&\quad \quad ~+\alpha (M_rN)^2\big)\big)\notag\\
&=\mathcal{O}\left(N_\text{epoch}N_\text{train}\alpha(I+(L-1)\alpha+1) (M_rN)^2\right),
\end{align}where $(a)$ comes from $f_l=\alpha M_rN$ for $1\leq l \leq L$. The constant $\alpha$ is to scale the hidden-layer nodes according to the number of antennas at the BS and UE.
In the prediction phase, the complexity is given by 
\begin{align}
C_\text{test}=\mathcal{O}\left(\alpha(I+(L-1)\alpha+1) (M_rN)^2\right).
\end{align}In addition, the complexity of LMMSE estimation in (\ref{g_n}) is $\mathcal{O}\left((M_rN)^3\right)$. 
Thus, the total complexity of MLP-based predictor is 
\begin{align}
C_\text{MLP}&=\mathcal{O}\big((N_\text{epoch}N_\text{train}+1)\alpha(I+(L-1)\alpha+1)(M_rN)^2\notag\\
&\quad \quad ~+(M_rN)^3\big).\label{complexity_MLP}
\end{align}The complexity of VKF-based predictor and MLP-based predictor is summarized in Table \ref{table1}.

To compare the total complexity of VKF-based predictor and MLP-based predictor, we approximate the complexity of both predictors as $\mathcal{O}\left((M_rN)^3\right)$ and $\mathcal{O}\left(N_\text{epoch}N_\text{train}(M_rN)^2\right)$. The MLP-based predictor has much higher complexity than the VKF-based predictor since $N_\text{epoch}N_\text{train} \gg M_rN$. However, once trained, the prediction phase complexity of MLP-based predictor becomes $\mathcal{O}\left((M_rN)^2\right)$, which is much lower than that of VKF-based predictor $\mathcal{O}\left((M_rN)^3\right)$. Thus, we can mitigate the high complexity problem of MLP-based predictor by conducting offline training.

\subsection{Numerical results}

In the simulation, the MLP is implemented with Deep Learning toolbox of MATLAB. For the MLP, the training rate is set to $0.001$ with the batch size $128$ and number of epochs $N_\text{epoch}=1000$. We consider the urban micro (UMi) cell in the SCM \cite{SCM}, where the mobility of UE is $3$ $\mathrm{km/h}$, the carrier frequency is $2.3$ GHz, and the time slot duration is $40$ ms. The BS is equipped with the $8\times8$ uniform planar array (UPA) at the height of $15$ m. Also, each UE is equipped with $N=2$ transmit antennas as the uniform linear array (ULA) and transmits the length $\tau=2$ pilot sequences. We adopt the pilot matrix $\boldsymbol{\Psi}_n$ using the discrete Fourier transform (DFT) matrix, which satisfies the assumption in Section \ref{subection:system model}. Note that the SCM takes the pathloss into account, resulting in very small channel gains. To fairly compare the channel predictors, we normalize the average gain of channel vectors to $M_r N$, i.e., $\mathbb{E}\left[\norm{\underline{\bh}_{n}}^2\right]=M_rN$, for all simulations. 
We define the normalized mean square error (NMSE) as the performance metric
\begin{align}
\text{NMSE}=\mathbb{E}\left[{\left\|\hat{\underline{\bh}}_{n+1}-\underline{\bh}_{n+1}\right\|}^2/\norm{\underline{\bh}_{n+1}}^2\right],
\end{align} where $\hat{\underline{\bh}}_{n+1}$ is the predicted channel, and $\underline{\bh}_{n+1}$ is the true channel.
The system parameters are summarized in Table \ref{table2} while we explicitly state parameter values if they are different from the table.

\begin{table}[t]
	\renewcommand{\arraystretch}{1.6}
	\captionsetup{justification=centering}
	\captionsetup{labelsep=newline}
	\caption{System parameters}
	\centering
	\label{table2}
	
	\begin{tabular}{l  l}
		\toprule
		Parameter & Value \\
		\midrule
		Environment & UMi\\
		
		Mobility of UE & {3} km/h\\
		
		Carrier frequency & 2.3 GHz\\
		
		Time slot duration  & {40 ms}\\
		
		BS antenna structure & 64, 8$\times$8 UPA\\
		\bottomrule
	\end{tabular}
\end{table}

\begin{figure}[t]
	\centering
	\includegraphics[width=9.4 cm]{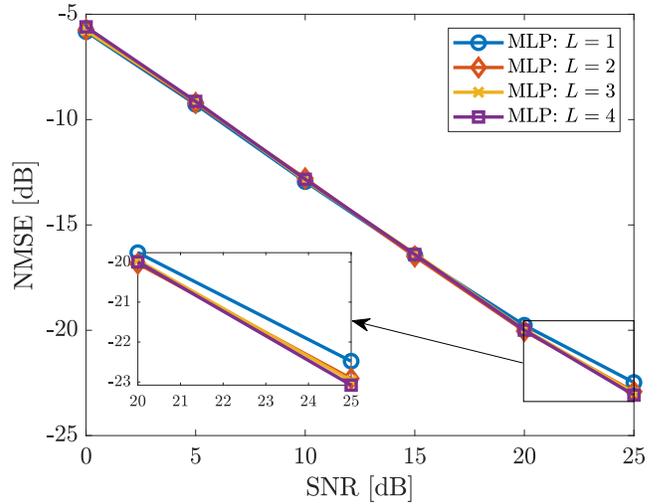}%
	\caption{NMSE of MLP-based predictor with different numbers of hidden-layer $L$ according to SNR with $I=3$, $N_{\text{train}}=2048$, and $f_l=512$.}
	\label{figure4}
\end{figure}

In our simulation, we compare the following methods:
\begin{itemize}
	\item \textbf{Outdated}: outdated channel $\hat{\underline{\bh}}_{n+1}^\text{outdated}=\boldsymbol{\bar{\Psi}}^{\dagger}\underline{\by}_n$ where $\boldsymbol{\bar{\Psi}}_n$ is the known pilot matrix. This serves as a baseline of channel predictors.
	\item \textbf{Extrapolation}: extrapolation-based prediction. We use the first-order polynomial-based extrapolation to predict the channel $\hat{\underline{\bh}}_{n+1}^{\textrm{ext}}=f_\textrm{ext}(n+1)=(n+1)\ba+\bb$ where the coefficient vectors $\ba$ and $\bb$ are determined by solving $\underline{\by}_n=n\ba+\bb$ and $\underline{\by}_{n-1}=(n-1)\ba+\bb$.
	\item \textbf{MLP without pre-processing}: MLP trained with the measurements $\big\{\underline{\by}_{n-I+1},\cdots,\underline{\by}_{n}\big\}$ without any pre-processing.
	\item \textbf{MLP with pre-processing (or simply MLP)}: MLP developed in Section \ref{Section:MLP}.
	\item \textbf{VKF}: vector Kalman filter-based prediction developed in Section \ref{Section:VKF}.
\end{itemize}

\noindent To have fair comparison, we set the number of measurement vectors $N_s$ in (\ref{N_s}) and number of training samples $N_\mathrm{train}$ in (\ref{N_train}) to be the same, i.e., $N_s=N_\mathrm{train}$.

\begin{figure}[t]
	\centering
	\includegraphics[width=9.4 cm]{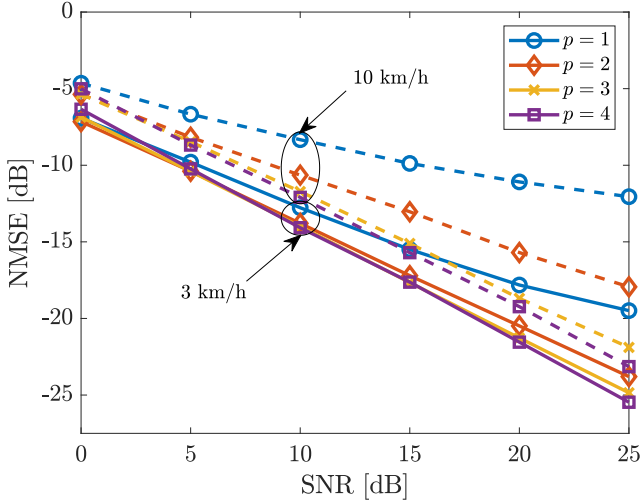}%
	\caption{NMSE of VKF-based predictor with different AR-order $p$ according to SNR with $N_{s}=2048$ and $v=3, 10$ km/h.}
	\label{figure5}
\end{figure}

\begin{figure}[t]
	\centering
	\includegraphics[width=9.4 cm]{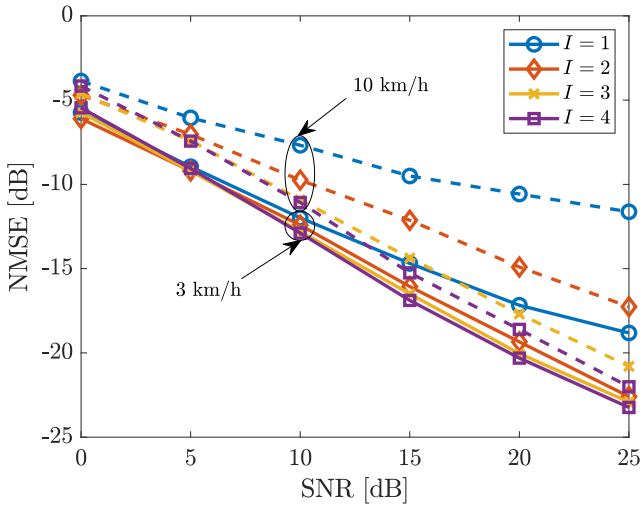}%
	\caption{NMSE of MLP-based predictor with different input-order $I$ according to SNR with $N_{\text{train}}=2048$ and $v=3, 10$ km/h.}
	\label{figure6}
\end{figure}

\begin{figure}[t]
	\centering
	\includegraphics[width=9.4 cm]{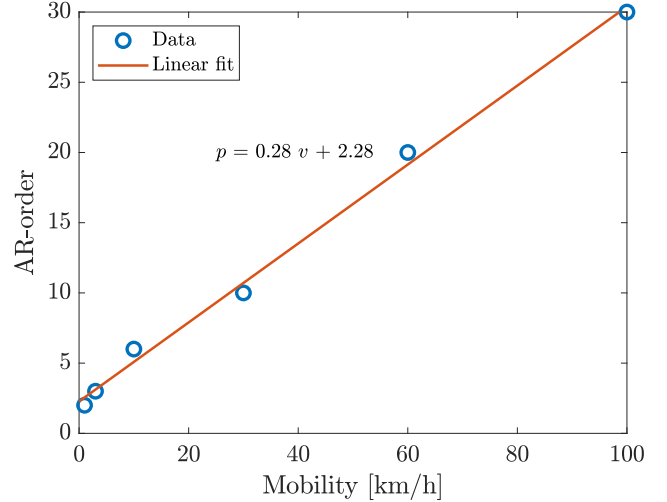}
	\caption{AR-order according to mobility of UE with $N_s=2048$.}
	\label{figure7}
\end{figure}

In Fig. \ref{figure4}, we compare the NMSE of MLP with different numbers of hidden-layer $L$ according to the SNR with the input-order $I=3$, $N_\text{train}=2048$, and $f_l=512$ for $1 \leq l \leq L$. It is obvious from the figure that $L=2$ is sufficient for channel prediction. Note that we do not use the activation layer (e.g. sigmoid or rectified linear unit (ReLU)) in the hidden-layer since it worsens the result. We set $L=2$ and $f_l=512$ for the following simulations to reduce the MLP training complexity.

Figs. \ref{figure5} and \ref{figure6} depict the NMSE of VKF-based predictor with different AR-order $p$ and that of MLP-based predictor with different input-order $I$ with $N_s=N_\text{train}=2048$ and $v=3, 10$ km/h. It is clear from the figures that higher order is needed for higher mobility to achieve the same prediction accuracy. Thus, it is crucial to find the proper order to balance the accuracy and complexity according to the mobility of UE. Since the AR-order and input-order balance the complexity and prediction accuracy, we use the same order for both predictors in the remaining simulations.

\begin{figure}[t]
	\centering
	\includegraphics[width=9.4 cm]{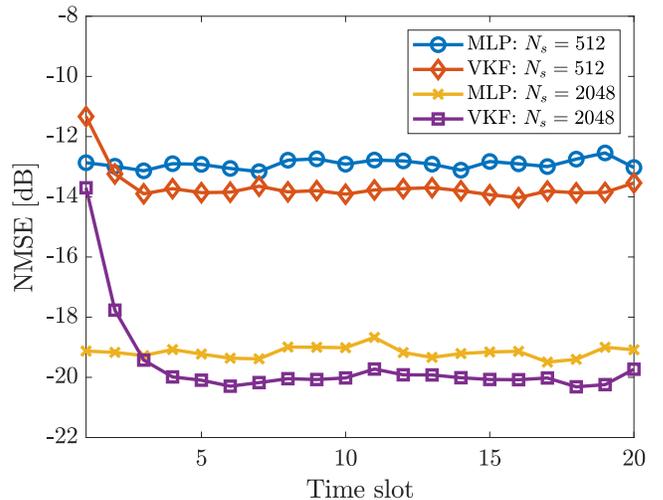}
	\caption{NMSE of VKF-based predictor and MLP-based predictor according to time slot with $N_s=512,~2048$, $p=I=3$, and $\text{SNR}=20$ dB.}
	\label{figure8}
\end{figure}

In Fig. \ref{figure7}, we numerically determine the effective AR-order according to the mobility of UE with $N_s=2048$. We define the effective order as the minimum AR-order satisfying $\textrm{NMSE}<-20$ dB with $\text{SNR}=20$ dB. Approximately, the relation $n_o=g(\hat{v})$ in (\ref{optimization}) turns out to be linear, i.e., $\text{AR-order}\approx 0.3\cdot \text{Mobility [km/h]}$. Thus, we can determine the effective complexity order by using the estimated mobility in Section \ref{Section:mobility}.

\begin{figure}[t]
	\centering
	\includegraphics[width=9.4 cm]{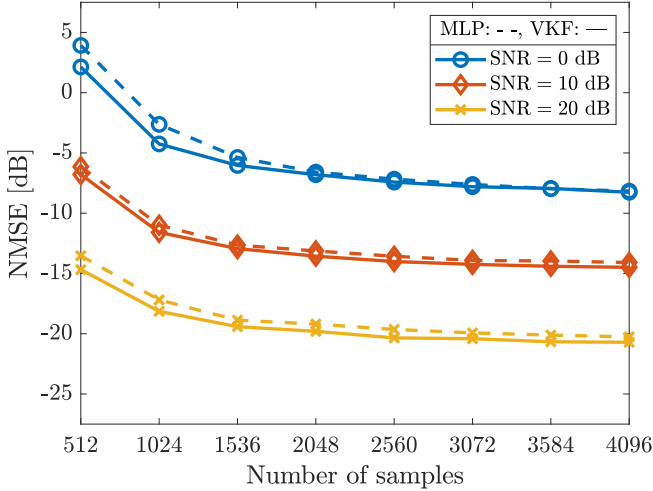}%
	\caption{NMSE of VKF-based predictor and MLP-based predictor with different SNR according to number of samples with $p=I=3$.}
	\label{figure9}
\end{figure}

\begin{figure}[t]
	\centering
	\includegraphics[width=9.4 cm]{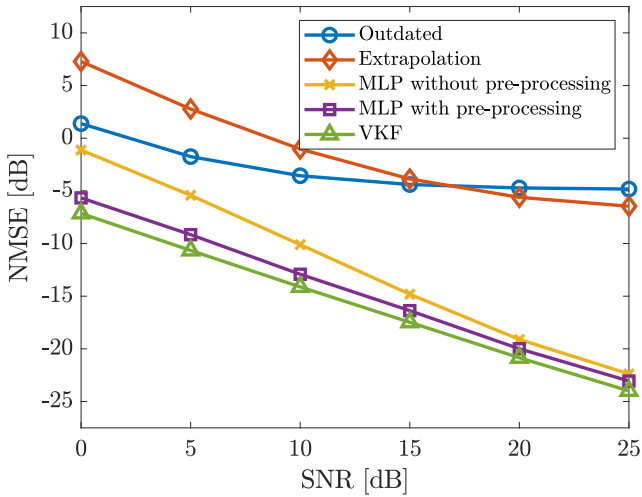}
	\caption{NMSE of outdated channel, extrapolation-based predictor, VKF-based predictor, and MLP-based predictor with and without pre-processing with respect to SNR with $p=I=3$ and $N_s=2048$.} 
	\label{figure10}
\end{figure}

Fig. \ref{figure8} shows the NMSE of VKF-based and MLP-based predictors according to the time slot with $N_s=512,~2048$, $p=I=3$, and $\text{SNR}=20$ dB. As the time slot increases, the VKF-based predictor outperforms the MLP-based predictor after only two successive predictions for both cases of $N_s$. Note that all the numerical results except those in Fig. \ref{figure8} are averaged over 100 time slots.

In Fig. \ref{figure9}, we verify the NMSE of VKF-based and MLP-based predictors with different SNR values according to the number of samples $N_s$ with $p=I=3$. Fig. \ref{figure9} shows that the VKF-based predictor requires less number of samples than the MLP-based predictor to achieve the same prediction accuracy for all SNR values.

\begin{figure}[t]
	\centering
	\includegraphics[width=9.4 cm]{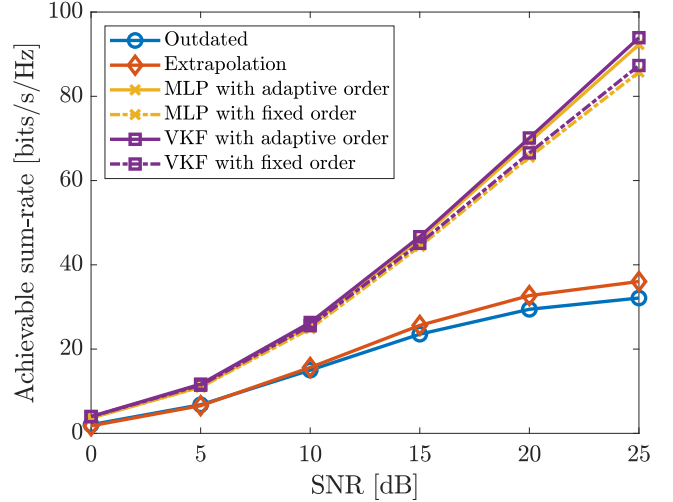}%
	\caption{Achievable sum-rate of outdated channel, extrapolation-based predictor, VKF-based predictor with adaptive order and fixed order $p=3$, and MLP-based predictor with adaptive order and fixed order $I=3$ according to SNR with $N_s=2048$ and $K=8$.}
	\label{figure11}
\end{figure}

We compare in Fig. \ref{figure10} the NMSE of outdated channel, extrapolation-based predictor, VKF-based predictor, and MLP-based predictor with and without pre-processing with respect to the SNR. We set $p=I=3$ and $N_s=2048$. The extrapolation-based predictor would not be able to track the SCM channel even in high SNR regime. The developed predictors outperform the outdated channel and extrapolation-based predictor. The MLP-based predictor with the pre-processing has almost the same performance as the VKF-based predictor. Also, the proposed LMMSE pre-processing gives about $5$ dB NMSE gain at $0$ dB SNR.

In Fig. \ref{figure11}, we compare the achievable sum-rates, defined in (\ref{achivable sum-rate}), of outdated channel, extrapolation-based predictor,  VKF-based predictor with adaptive order and fixed order $p=3$, and MLP-based predictor with adaptive order and fixed order $I=3$ according to the SNR with $N_s=2048$ and number of UEs $K=8$. The first four UEs experience the mobility $v=3$ km/h while the other UEs experience the mobility $v=10$ km/h with different geometries. We adaptively select the order with respect to the mobility of UE as in Fig. \ref{figure7}. The achievable sum-rates of outdated channel and extrapolation-based predictor increase as the SNR increases, but saturate in high SNR regime. The VKF-based predictor and MLP-based predictor have substantial gain over the extrapolation-based predictor, and the gain becomes larger as the SNR increases. Both predictors have almost the same sum-rates since we set the same complexity order for both predictors. Also, the adaptive order selection provides additional gain for both predictors. 

\section{Conclusion}\label{Section:conclusion}
In this paper, we developed the channel predictors for the time-varying massive MIMO systems. First, we implemented the low-complexity mobility estimator to set proper prediction complexity order. Then, we developed the VKF-based predictor with the AR parameter estimation from the SCM data. We also developed the MLP-based channel predictor based on the LMMSE noise pre-processing. The numerical results showed that both channel predictors have substantial gain over the outdated channel in terms of the prediction accuracy and sum-rate. Regarding the complexity, the MLP-based predictor has much higher total complexity than the VKF-based predictor. Once trained, however, the MLP-based predictor has much lower complexity, which shows offline learning is crucial to adopt the MLP-based channel predictor in practice. It might be also possible to exploit more advanced ML techniques, e.g., meta learning or few-shot learning, to mitigate the training overhead, which is an interesting future research topic. We believe the complexity analysis and numerical results in this paper can give good guidelines for system operators to adopt a better channel prediction technique depending on their situation, e.g., whether it is possible to have a large number of training samples or to perform the offline learning.

\ifCLASSOPTIONcaptionsoff
  \newpage
\fi



\bibliographystyle{IEEEtran}
\bibliography{IEEEtcom}

\begin{thebibliography}{10}
\providecommand{\url}[1]{#1}
\csname url@samestyle\endcsname
\providecommand{\newblock}{\relax}
\providecommand{\bibinfo}[2]{#2}
\providecommand{\BIBentrySTDinterwordspacing}{\spaceskip=0pt\relax}
\providecommand{\BIBentryALTinterwordstretchfactor}{4}
\providecommand{\BIBentryALTinterwordspacing}{\spaceskip=\fontdimen2\font plus
\BIBentryALTinterwordstretchfactor\fontdimen3\font minus
  \fontdimen4\font\relax}
\providecommand{\BIBforeignlanguage}[2]{{%
\expandafter\ifx\csname l@#1\endcsname\relax
\typeout{** WARNING: IEEEtran.bst: No hyphenation pattern has been}%
\typeout{** loaded for the language `#1'. Using the pattern for}%
\typeout{** the default language instead.}%
\else
\language=\csname l@#1\endcsname
\fi
#2}}
\providecommand{\BIBdecl}{\relax}
\BIBdecl

\bibitem{Marzetta06}
T.~L. Marzetta, ``Noncooperative {c}ellular {w}ireless with {u}nlimited
  {n}umbers of {b}ase {s}tation {a}ntennas,'' \emph{IEEE Transactions on
  Wireless Communications}, vol.~9, no.~11, pp. 3590--3600, Nov. 2010.

\bibitem{Papa17}
A.~K. {Papazafeiropoulos}, ``Impact of general channel aging conditions on the
  downlink performance of massive {MIMO},'' \emph{IEEE Transactions on
  Vehicular Technology}, vol.~66, no.~2, pp. 1428--1442, Feb. 2017.

\bibitem{6608213}
K.~T. {Truong} and R.~W. {Heath}, ``Effects of channel aging in massive {MIMO}
  systems,'' \emph{Journal of Communications and Networks}, vol.~15, no.~4, pp.
  338--351, Aug. 2013.

\bibitem{5947055}
N.~{Palleit} and T.~{Weber}, ``Time prediction of non flat fading channels,''
  in \emph{2011 IEEE International Conference on Acoustics, Speech and Signal
  Processing (ICASSP)}, May 2011, pp. 2752--2755.

\bibitem{Kong15}
C.~{Kong}, C.~{Zhong}, A.~K. {Papazafeiropoulos}, M.~{Matthaiou}, and
  Z.~{Zhang}, ``Sum-rate and power scaling of massive {MIMO} systems with
  channel aging,'' \emph{IEEE Transactions on Communications}, vol.~63, no.~12,
  pp. 4879--4893, Dec. 2015.

\bibitem{6815892}
V.~{Jungnickel}, K.~{Manolakis}, W.~{Zirwas}, B.~{Panzner}, V.~{Braun},
  M.~{Lossow}, M.~{Sternad}, R.~{Apelfröjd}, and T.~{Svensson}, ``The role of
  small cells, coordinated multipoint, and massive {MIMO} in {5G},'' \emph{IEEE
  Communications Magazine}, vol.~52, no.~5, pp. 44--51, May 2014.

\bibitem{8724442}
V.~{Arya} and K.~{Appaiah}, ``Kalman filter based tracking for channel aging in
  massive {MIMO} systems,'' in \emph{2018 International Conference on Signal
  Processing and Communications (SPCOM)}, Jul. 2018, pp. 362--366.

\bibitem{6823657}
S.~{Noh}, M.~D. {Zoltowski}, Y.~{Sung}, and D.~J. {Love}, ``Pilot beam pattern
  design for channel estimation in massive {MIMO} systems,'' \emph{IEEE Journal
  of Selected Topics in Signal Processing}, vol.~8, no.~5, pp. 787--801, Oct.
  2014.

\bibitem{Choi142}
J.~{Choi}, D.~J. {Love}, and P.~{Bidigare}, ``Downlink training techniques for
  {FDD} massive {MIMO} systems: Open-loop and closed-loop training with
  memory,'' \emph{IEEE Journal of Selected Topics in Signal Processing},
  vol.~8, no.~5, pp. 802--814, Oct. 2014.

\bibitem{7938362}
S.~{Bazzi} and W.~{Xu}, ``Downlink training sequence design for {FDD} multiuser
  massive {MIMO} systems,'' \emph{IEEE Transactions on Signal Processing},
  vol.~65, no.~18, pp. 4732--4744, Sep. 2017.

\bibitem{Kim2019}
H.~Kim and J.~Choi, ``Channel estimation for spatially/temporally correlated
  massive {MIMO} systems with one-bit {ADC}s,'' \emph{EURASIP Journal on
  Wireless Communications and Networking}, vol. 2019, no.~1, p. 267, Dec. 2019.

\bibitem{7511214}
C.~{Li}, J.~{Zhang}, S.~{Song}, and K.~B. {Letaief}, ``Selective uplink
  training for massive {MIMO} systems,'' in \emph{2016 IEEE International
  Conference on Communications (ICC)}, May 2016, pp. 1--6.

\bibitem{Haykin96}
S.~Haykin, \emph{Adaptive Filter Theory (3rd Ed.)}.\hskip 1em plus 0.5em minus
  0.4em\relax Upper Saddle River, NJ, USA: Prentice-Hall, Inc., 1996.

\bibitem{Baddour05}
K.~E. {Baddour} and N.~C. {Beaulieu}, ``Autoregressive modeling for fading
  channel simulation,'' \emph{IEEE Transactions on Wireless Communications},
  vol.~4, no.~4, pp. 1650--1662, Jul. 2005.

\bibitem{Kashyap17}
S.~{Kashyap}, C.~{Mollén}, E.~{Björnson}, and E.~G. {Larsson}, ``Performance
  analysis of {TDD} massive {MIMO} with {K}alman channel prediction,'' in
  \emph{2017 IEEE International Conference on Acoustics, Speech and Signal
  Processing (ICASSP)}, Mar. 2017, pp. 3554--3558.

\bibitem{Shikur15}
B.~Y. {Shikur} and T.~{Weber}, ``Channel prediction using an adaptive {K}alman
  filter,'' in \emph{WSA 2015; 19th International ITG Workshop on Smart
  Antennas}, Mar. 2015, pp. 1--7.

\bibitem{Wang17}
T.~{Wang}, C.~{Wen}, H.~{Wang}, F.~{Gao}, T.~{Jiang}, and S.~{Jin}, ``Deep
  learning for wireless physical layer: Opportunities and challenges,''
  \emph{China Communications}, vol.~14, no.~11, pp. 92--111, Nov. 2017.

\bibitem{Wen18}
C.~{Wen}, W.~{Shih}, and S.~{Jin}, ``Deep learning for massive {MIMO} {CSI}
  feedback,'' \emph{IEEE Wireless Communications Letters}, vol.~7, no.~5, pp.
  748--751, Oct. 2018.

\bibitem{Wang19}
T.~{Wang}, C.~{Wen}, S.~{Jin}, and G.~Y. {Li}, ``Deep learning-based {CSI}
  feedback approach for time-varying massive {MIMO} channels,'' \emph{IEEE
  Wireless Communications Letters}, vol.~8, no.~2, pp. 416--419, Apr. 2019.

\bibitem{Dong19}
P.~{Dong}, H.~{Zhang}, G.~Y. {Li}, N.~{NaderiAlizadeh}, and I.~S. {Gaspar},
  ``Deep {CNN} for wideband mmwave massive {MIMO} channel estimation using
  frequency correlation,'' in \emph{2019 IEEE International Conference on
  Acoustics, Speech and Signal Processing (ICASSP)}, May 2019, pp. 4529--4533.

\bibitem{Soltani19}
M.~{Soltani}, V.~{Pourahmadi}, A.~{Mirzaei}, and H.~{Sheikhzadeh}, ``Deep
  learning-based channel estimation,'' \emph{IEEE Communications Letters},
  vol.~23, no.~4, pp. 652--655, Apr. 2019.

\bibitem{8815557}
J.~{Yuan}, H.~Q. {Ngo}, and M.~{Matthaiou}, ``Machine learning-based channel
  estimation in massive {MIMO} with channel aging,'' in \emph{2019 IEEE 20th
  International Workshop on Signal Processing Advances in Wireless
  Communications (SPAWC)}, Jul. 2019, pp. 1--5.

\bibitem{Krasny01}
L.~{Krasny}, H.~{Arslan}, D.~{Koilpillai}, and S.~{Chennakeshu}, ``Doppler
  spread estimation in mobile radio systems,'' \emph{IEEE Communications
  Letters}, vol.~5, no.~5, pp. 197--199, May 2001.

\bibitem{Baddour051}
K.~E. {Baddour} and N.~C. {Beaulieu}, ``Robust {D}oppler spread estimation in
  nonisotropic fading channels,'' \emph{IEEE Transactions on Wireless
  Communications}, vol.~4, no.~6, pp. 2677--2682, Nov. 2005.

\bibitem{Zheng09}
Y.~R. {Zheng} and C.~{Xiao}, ``Mobile speed estimation for broadband wireless
  communications over {R}ician fading channels,'' \emph{IEEE Transactions on
  Wireless Communications}, vol.~8, no.~1, pp. 1--5, Jan. 2009.

\bibitem{SCM}
\emph{Study on 3{D} channel model for {LTE}}, 3GPP TR 36.873 V12.7.0 Std., Jan.
  2018.

\bibitem{Pan07}
S.~{Pan}, S.~{Durrani}, and M.~{Bialkowski}, ``{MIMO} capacity for spatial
  channel model scenarios,'' in \emph{IEEE Proceedings Australian Communication
  Theory Workshop Adelade Australia}, Feb. 2007, pp. 25--29.

\bibitem{6632104}
Y.~{Liu} and L.~{Li}, ``Adaptive multi-step channel prediction in spatial
  channel model using {K}alman filter,'' in \emph{ICT 2013}, May 2013, pp.
  1--5.

\bibitem{durbin60}
J.~Durbin, ``The fitting of time-series models,'' \emph{Revue de l'Institut
  International de Statistique / Review of the International Statistical
  Institute}, vol.~28, no.~3, pp. 233--244, 1960.

\bibitem{Hung10}
K.~{Hung} and D.~W. {Lin}, ``Pilot-based {LMMSE} channel estimation for {OFDM}
  systems with power–delay profile approximation,'' \emph{IEEE Transactions
  on Vehicular Technology}, vol.~59, no.~1, pp. 150--159, Jan. 2010.

\bibitem{8353153}
H.~{He}, C.~{Wen}, S.~{Jin}, and G.~Y. {Li}, ``Deep learning-based channel
  estimation for beamspace mmwave massive {MIMO} systems,'' \emph{IEEE Wireless
  Communications Letters}, vol.~7, no.~5, pp. 852--855, Oct. 2018.

\bibitem{8651830}
S.~{Khan}, K.~S. {Khan}, and S.~Y. {Shin}, ``Symbol denoising in high order
  {M-QAM} using residual learning of deep {CNN},'' in \emph{2019 16th IEEE
  Annual Consumer Communications Networking Conference (CCNC)}, Jan. 2019, pp.
  1--6.

\bibitem{hunger2005floating}
R.~Hunger, \emph{Floating point operations in matrix-vector calculus}.\hskip
  1em plus 0.5em minus 0.4em\relax Munich University of Technology, Inst. for
  Circuit Theory and Signal, 2005.

\bibitem{939044}
E.~{Mizutani} and S.~E. {Dreyfus}, ``On complexity analysis of supervised
  {MLP}-learning for algorithmic comparisons,'' in \emph{IJCNN'01.
  International Joint Conference on Neural Networks. Proceedings (Cat.
  No.01CH37222)}, vol.~1, Jul. 2001, pp. 347--352.

\end{thebibliography}
%
%



%




\end{document}